\documentclass[11pt]{article}

\usepackage[hyphens]{url}
\usepackage[pagebackref,colorlinks]{hyperref}
\usepackage[hyphenbreaks]{breakurl}
\usepackage{amsmath} 
\usepackage{amsthm} 
\usepackage{amssymb}	
\usepackage{graphicx} 
\usepackage{multicol} 
\usepackage{multirow}
\usepackage{color}
\usepackage[dvips,letterpaper,margin=1in,bottom=1in]{geometry}
\usepackage[capitalize,noabbrev]{cleveref}

\usepackage{bm}
\usepackage{bbm}

\usepackage{ytableau}
\usepackage{longtable}

\usepackage[utf8]{inputenc}
\usepackage[english]{babel}

\usepackage[T1]{fontenc}
\AtBeginDocument{%
  \DeclareFontShape{T1}{cmr}{m}{scit}{<->ssub*cmr/m/sc}{}%
}

\usepackage{diagbox}
\usepackage{mathtools}
\allowdisplaybreaks

\newtheorem{theorem}{Theorem}[section]

\newtheorem{lemma}[theorem]{Lemma}

\newtheorem{proposition}[theorem]{Proposition}

\newtheorem{definition}[theorem]{Definition}

\newtheorem{remark}{Remark}[section]

\DeclarePairedDelimiter\rbra{\lparen}{\rparen}
\DeclarePairedDelimiter\sbra{\lbrack}{\rbrack}
\DeclarePairedDelimiter\cbra{\{}{\}}
\DeclarePairedDelimiter\abs{\lvert}{\rvert}
\DeclarePairedDelimiter\Abs{\lVert}{\rVert}
\DeclarePairedDelimiter\ceil{\lceil}{\rceil}

\DeclarePairedDelimiter\ket{\lvert}{\rangle}
\DeclarePairedDelimiter\bra{\langle}{\rvert}

\DeclareMathOperator*{\E}{\mathbb{E}}

\newcommand{\tr} {\operatorname{tr}}

\newcommand{\ketbra}[2]{\ensuremath{\ket{#1}\!\bra{#2}}}

\usepackage{latexsym}
\usepackage{CJK}

\usepackage{enumerate}
\usepackage{enumitem}

\usepackage{algorithm}
\usepackage{algpseudocode}

\usepackage{stmaryrd}
\usepackage{tabularx}
\usepackage{booktabs}
\usepackage{adjustbox}

\newcommand{\footremember}[2]{%
    \footnote{#2}
    \newcounter{#1}
    \setcounter{#1}{\value{footnote}}%
}

\usepackage{tikz}

\usetikzlibrary{quantikz2}

\begin{document}

\title{Optimal Min-Entropy Estimation with High Probability}
\author{Qisheng Wang \footremember{1}{Qisheng Wang is with the School of Computer Science, Shanghai Jiao Tong University (e-mail: \url{QishengWang1994@gmail.com}).}}
\date{}

\maketitle

\begin{abstract}
    Min-entropy plays a fundamental role in assessing the unpredictability of randomness sources. 
    In particular, NIST SP 800-90B specifies min-entropy estimation as a key procedure for entropy sources used in cryptographic random bit generators, illustrating the practical importance of reliable min-entropy estimation. 
    In this paper, we settle the sample complexity of min-entropy estimation with high probability in both the classical and quantum cases. 
    Specifically, for min-entropy estimation with additive error $\varepsilon$ and probability at least $1-\delta$, the sample complexity is $\Theta(d\log(d/\delta)/\varepsilon^2)$ 
    for classical distributions on $d$ symbols and $\Theta((d^2+d\log(1/\delta))/\varepsilon^2)$ 
    for quantum states on a $d$-dimensional Hilbert space. 
    Our results admit short and simple proofs.
\end{abstract}



\section{Introduction}

The min-entropy of a discrete probability distribution $p$ on $\sbra{d} = \cbra{1, 2, \dots, d}$, defined by 
\[
\mathrm{H}_\infty\rbra{p} = -\log\rbra*{ \max_{i \in \sbra{d}} p_i },
\]
as the limiting case of the R\'enyi entropy \cite{Ren61} of order $\alpha = \infty$, quantifies the worst-case unpredictability of a random variable \cite{KRS09} and plays a fundamental role in randomness extraction \cite{IZ89,BBCM95}. 
In particular, the Leftover Hash Lemma \cite{ILL89,BDK+11} (cf.\ \cite[Theorem 6.18]{Vad12}) shows that a source with $k$ bits of min-entropy admits a strong extractor that produces $k - 2\log_2\rbra{1/\eta}$ nearly uniform bits with error $\eta$ in the total variation distance. 

The estimation of min-entropy has been utilized as a key procedure in assessing entropy sources used for random bit generators in NIST SP 800-90B \cite{TBK+18}. 
In particular, the National Institute of Standards and Technology (NIST) Cryptographic Module Validation Program (CMVP) requires applicable FIPS 140-2 and FIPS 140-3 validation submissions to include documentation justifying conformance to NIST SP 800-90B (cf.\ \cite[Sections 7.18–7.19]{FIPS140-2} and \cite[Section 4.8]{FIPS140-3}). 
However, the sample complexity of min-entropy estimation with high precision and probability remains open. 
Recently, a min-entropy estimator was provided in \cite{AK26} with sample complexity $O\rbra{{d \log\rbra{d/\delta}}/{\varepsilon^2}}$ with $\varepsilon$ the additive error and $\delta$ the error probability, together with a lower bound of $\Omega\rbra{d\log\rbra{d}}$.\footnote{Previously, the sample complexity of min-entropy estimation was claimed as $\Theta\rbra{d/\log\rbra{d}}$ in \cite{LW19} by mistake.} 
In this paper, we close the sample complexity gap by establishing a tight lower bound.
In addition, we also provide a streamlined proof of the upper bound. 
Both of our proofs are substantially shorter, compared to the proofs in \cite[Sections~7 and~4]{AK26} each of which is eight pages long.

\begin{theorem}[Optimal min-entropy estimation] \label{thm:classical-main}
    The sample complexity of estimating the min-entropy of an unknown distribution on $\sbra{d}$ to within additive error $\varepsilon$ with success probability $1 - \delta$ is 
    \[
    \Theta\!\left(\frac{d \log(d/\delta)}{\varepsilon^2}\right).
    \]
\end{theorem}

We further investigate the quantum analog of min-entropy estimation. 
For a quantum state $\rho$, its min-entropy, as a quantum generalization of classical min-entropy, is defined by
\[
\mathrm{S}_\infty\rbra{\rho} = -\log\rbra*{\Abs{\rho}_\infty}, 
\]
where $\Abs{\cdot}_\infty$ denotes the operator norm.
It is a fundamental information measure with operational meanings in quantum information theory \cite{Dat09,KRS09,LCJ21}. 
In particular, the min-entropy of a reduced state exactly determines the number of Bell pairs obtainable through deterministic entanglement concentration \cite{MK01}. 
Recently, min-entropy has been shown to govern fundamental limitations on entanglement distillation with limited computational resources \cite{LREJ25}. 
These operational interpretations naturally raise the question of how min-entropy can be estimated when the underlying quantum state is unknown and accessible only through independent copies.

The estimation of quantum min-entropy was discussed in \cite{WZ25}: for $d$-dimensional quantum states and additive error $\varepsilon$, a quantum min-entropy estimator with sample complexity $O\rbra{d^2/\varepsilon^2}$ is implied by \cite[Theorem 1.18]{OW17}, whereas the prior best lower bound is $\Omega\rbra{d/\varepsilon + d\log\rbra{d}}$ by combining \cite[Theorem VI.4]{WZ25} and \cite[Theorem 7.1]{AK26}.
For success probability $1 - \delta$, the estimator of \cite{OW17} implies an estimator with sample complexity $O\rbra{d^2\log\rbra{1/\delta}/\varepsilon^2}$ using the median trick. 
An immediate question is whether the median trick already yields optimal quantum min-entropy estimation with high probability. 
In this paper, we show that weak Schur sampling \cite{CHW07} can beat the median trick and establish its optimality through a matching lower bound. 

\begin{theorem}[Optimal quantum min-entropy estimation] \label{thm:quantum-main}
    The sample complexity of estimating the min-entropy of an unknown $d$-dimensional quantum state to within additive error $\varepsilon$ with success probability $1 - \delta$ is
    \[
    \Theta\!\left(\frac{d^2+d\log\rbra{1/\delta}}{\varepsilon^2}\right).
    \]
\end{theorem}

\cref{thm:quantum-main} shows that the quantum min-entropy estimator implied by \cite{OW17} is optimal for a constant error $\delta$, say $\delta = 1/3$. 

\begin{remark}[Property testing with high probability]
We have established tight sample complexity bounds for \textit{high probability} entropy estimation in both the classical and quantum cases, which add to the high probability property testing literature. 
To our knowledge, tight bounds in the high probability regime are only known for identity/closeness/independence testing \cite{DGPP18,DGK+21} and integer-order R\'enyi entropy estimation \cite{Sko26}.
\end{remark}

\section{Classical Min-Entropy Estimation}

\subsection{Upper Bounds}

Suppose that $p$ is an unknown probability distribution on $\sbra{d}$. 
Let $\bm{X} = \rbra{X_1, X_2, \dots, X_m}$ be a random sample from $p^{\otimes m}$. 
Define the estimator 
\[
\widehat{H} = -\log\rbra*{\max_{i \in \sbra{d}} \frac{N_i}{m}}, \qquad N_i = \sum_{j=1}^m \mathbf{1}_{\cbra{X_j = i}}.
\]

\begin{theorem} \label{thm:classical-upper}
    It holds that $\Pr\sbra{\abs{\widehat{H} - \mathrm{H}_\infty\rbra{p}} \leq \varepsilon} \geq 1 - \delta$ for $\varepsilon, \delta \in \rbra{0, 1}$ by setting 
    \[
    m = \ceil*{ \frac{12d}{\varepsilon^2}\log\rbra*{\frac{d+1}{\delta}} } = O\rbra*{\frac{d}{\varepsilon^2}\log\rbra*{\frac{d}{\delta}}}.
    \]
\end{theorem}

To prove \cref{thm:classical-upper}, it suffices to show the following property. 

\begin{proposition} \label{prop:Pr-leq}
    Let $\hat p = \max_{i \in \sbra{d}} N_i/m$ and $p_\star = \max_{i\in\sbra{d}} p_i$. 
    Then, 
    \[
    \Pr\sbra*{ \abs*{\hat p - p_\star} > \frac{\varepsilon p_\star}{2} } \leq \rbra{d+1} \exp\rbra*{-\frac{m\varepsilon^2}{12d}}.
    \]
\end{proposition}

To prove \cref{prop:Pr-leq}, we need the Chernoff--Hoeffding inequality of the following form. 

\begin{lemma}[Chernoff--Hoeffding, cf.\ {\cite[Theorem 1.1]{DP09}}] \label{lemma:Chernoff}
    Let $X = \sum_{i=1}^n X_i$ with $X_i$ independently distributed in $\sbra{0, 1}$. 
    Then, for $\varepsilon \in \rbra{0, 1}$, 
    \begin{align*}
        \Pr\sbra*{ X > \rbra{1+\varepsilon} \E\sbra{X} } & \leq \exp\rbra*{-\frac{\varepsilon^2}{3}\E\sbra{X}}, \\
        \Pr\sbra*{ X < \rbra{1-\varepsilon} \E\sbra{X} } & \leq \exp\rbra*{-\frac{\varepsilon^2}{2}\E\sbra{X}}.
    \end{align*}
\end{lemma}

\begin{proof}[Proof of \cref{prop:Pr-leq}]
    We split the probability into two terms:
    \begin{equation} \label{eq:Pr-split}
    \Pr\sbra*{ \abs*{\hat p - p_\star} > \frac{\varepsilon p_\star}{2} } \leq \Pr\sbra*{\hat p < \rbra*{1 - \frac{\varepsilon}{2}} p_\star} + \Pr\sbra*{\hat p > \rbra*{1 + \frac{\varepsilon}{2}} p_\star}. 
    \end{equation}
    For the first term, suppose $p_{i_\star} = p_\star$ for some $i_\star \in \sbra{d}$. 
    Then, $\hat p \geq N_{i_\star} / m$.
    By \cref{lemma:Chernoff} and the fact that $\E\sbra{N_{i_\star}} = mp_\star$, 
    \begin{align} \label{eq:Pr-term1}
    \Pr\sbra*{\hat p < \rbra*{1 - \frac{\varepsilon}{2}} p_\star} \leq \Pr\sbra*{N_{i_\star} < \rbra*{1 - \frac{\varepsilon}{2}} mp_\star} \leq \exp\rbra*{-\frac{\varepsilon^2}{8} mp_\star}.
    \end{align}
    For the second term, by the union bound and \cref{lemma:Chernoff}, 
    \begin{align}
    \Pr\sbra*{\hat p > \rbra*{1 + \frac{\varepsilon}{2}} p_\star} 
    & \leq \sum_{i=1}^d \Pr\sbra*{ N_i > \rbra*{1 + \frac{\varepsilon}{2}} mp_\star } \nonumber \\
    & \leq d \Pr_{Z \sim \textup{Bin}\rbra{m, p_\star}}\sbra*{ Z > \rbra*{1 + \frac{\varepsilon}{2}} mp_\star } \nonumber \\
    & \leq d \exp\rbra*{-\frac{\varepsilon^2}{12}mp_\star}. \label{eq:Pr-term2}
    \end{align}
    Combining \cref{eq:Pr-split,eq:Pr-term1,eq:Pr-term2} gives 
    \[
    \Pr\sbra*{ \abs*{\hat p - p_\star} > \frac{\varepsilon p_\star}{2} } \leq \exp\rbra*{-\frac{\varepsilon^2}{8} mp_\star} + d \exp\rbra*{-\frac{\varepsilon^2}{12}mp_\star}.
    \]
    The proof is completed by noting that $p_\star \geq 1/d$. 
\end{proof}

\subsection{Lower Bounds}

\begin{theorem} \label{thm:clower}
    Any estimator that estimates the min-entropy of a distribution $p$ on $\sbra{d}$ to within additive error $\varepsilon$ with success probability at least $1 - \delta$ requires sample complexity at least 
    \[
    \Omega\rbra*{\frac{d}{\varepsilon^2}\log\rbra*{\frac{d}{\delta}}}.
    \]
\end{theorem}

\begin{proof}
    Let $u = \rbra{\frac{1}{d}, \dots, \frac{1}{d}}$ be the uniform distribution. 
    For each $j \in \sbra{d}$, define the distribution $q^{\rbra{j}}$ as
    \[
    q_j^{\rbra{j}} = \frac{1+4\varepsilon}{d}, \qquad q_i^{\rbra{j}} = \frac{1-4\varepsilon/\rbra{d-1}}{d} ~ (i \neq j). 
    \]
    It holds that $\Delta \coloneqq \mathrm{H}_\infty\rbra{u} - \mathrm{H}_\infty\rbra{q^{\rbra{j}}} = \log\rbra{1+4\varepsilon} > 2\varepsilon$.
    
    Assume that there is an estimator $\widehat{H}$ with sample complexity $m$ that estimates the min-entropy to within additive error $\varepsilon$ with success probability at least $1-\delta$.
    That is, for any distribution $p$ and a sample $\bm{X} = \rbra{X_1, X_2, \dots, X_m}$ from $p^{\otimes m}$, $\Pr\sbra{\abs{\widehat{H}\rbra{\bm{X}} - \mathrm{H}_\infty\rbra{p}} \leq \varepsilon} \geq 1 - \delta$. 
    Then, the estimator implies a tester that distinguishes $u^{\otimes m}$ from the average distribution 
    \[
    P = \frac{1}{d} \sum_{j=1}^d \rbra{q^{\rbra{j}}}^{\otimes m}, \qquad P\rbra{\bm{x}} = \frac{1}{d} \sum_{j=1}^d \prod_{t=1}^m q^{\rbra{j}}_{x_t},
    \]
    with success probability at least $1-\delta$, which return $u^{\otimes m}$ if $\widehat{H}\rbra{\bm{X}} > \log\rbra{d} - \Delta/2$ and $P$ otherwise. 
    To see the correctness, note that: (i) a sample $\bm{X}$ from $u^{\otimes m}$ satisfies $\Pr\sbra{\abs{ \widehat{H}\rbra{\bm{X}} - \mathrm{H}_\infty\rbra{u} } \leq \varepsilon} \geq 1 - \delta$, and (ii) a sample $\bm{X}$ from $P$ satisfies $\Pr\sbra{\abs{ \widehat{H}\rbra{\bm{X}} - \mathrm{H}_\infty\rbra{q^{\rbra{1}}} } \leq \varepsilon} \geq 1 - \delta$. 
    For $\bm{x} \in \sbra{d}^m$, define $f\rbra{\bm{x}} = \Pr\sbra{ \widehat{H}\rbra{\bm{x}} \leq \log\rbra{d} - \Delta/2 }$,
    where the probability may come from the randomness in the definition of $\widehat{H}$.
    Then, (i) gives that 
    \[
    \E_{\bm{X} \sim u^{\otimes m}}\sbra{f\rbra{\bm{X}}}  \leq \delta,
    \]
    and (ii) gives that
    \[
    1 - \delta \leq \E_{\bm{X} \sim P} \sbra*{f\rbra{\bm{X}}} = \sum_{\bm{x} \in \sbra{d}^m} P\rbra{\bm{x}} f\rbra{\bm{x}} = d^m \E_{\bm{X} \sim u^{\otimes m}} \sbra*{ P\rbra{\bm{X}} f\rbra{\bm{X}} }.
    \]
    By the Cauchy--Schwarz inequality and the above two inequalities, 
    \begin{equation} \label{eq:EPX-lb}
    \E_{\bm{X} \sim u^{\otimes m}} \sbra*{ P\rbra{\bm{X}}^2 } \geq \frac{\rbra*{\E_{\bm{X} \sim u^{\otimes m}} \sbra*{ P\rbra{\bm{X}} f\rbra{\bm{X}} }}^2}{\E_{\bm{X} \sim u^{\otimes m}}\sbra{f\rbra{\bm{X}}^2}} \geq \frac{\rbra{1-\delta}^2}{d^{2m} \delta}.
    \end{equation}
    On the other hand, direct calculation gives
    \begin{align}
    \E_{\bm{X} \sim u^{\otimes m}} \sbra*{ P\rbra{\bm{X}}^2 } 
    & = \frac{1}{d^{2m+1}} \rbra*{ 1 + \frac{16\varepsilon^2}{d-1} }^m + \frac{d-1}{d^{2m+1}} \rbra*{ 1 - \frac{16\varepsilon^2}{\rbra{d-1}^2} }^m \nonumber \\
    & \leq \frac{1}{d^{2m}} \rbra*{ 1 + \frac{e^{16m\varepsilon^2/\rbra{d-1}} - 1}{d} }. \label{eq:EPX-ub}
    \end{align}
    Combining \cref{eq:EPX-lb,eq:EPX-ub} gives 
    \[
    m \geq \frac{d-1}{16\varepsilon^2} \log\rbra*{ 1 + \frac{d\rbra{1-3\delta+\delta^2}}{\delta} } = \Omega\rbra*{\frac{d}{\varepsilon^2}\log\rbra*{\frac{d}{\delta}}}.
    \]
\end{proof}

\section{Quantum Min-Entropy Estimation}

\subsection{Upper Bounds}

Our estimator employs the weak Schur sampling \cite{CHW07}. 
Here, we use the formulation in \cite{OW21}.

\begin{definition}[Weak Schur sampling]
    Let $\rho \in \mathbb{C}^{d \times d}$ be a density operator. 
    By the Schur--Weyl duality, 
    \[
    \rbra{\mathbb{C}^{d}}^{\otimes n} \cong \bigoplus_{\substack{\lambda \vdash n \\ \ell\rbra{\lambda} \leq d}} \mathcal{P}_\lambda \otimes \mathcal{Q}_\lambda^d,
    \]
    where $\mathcal{P}_\lambda$ and $\mathcal{Q}_\lambda^d$ are the irreducible representations of the symmetric group $\mathbb{S}_n$ and the unitary group $\mathbb{U}_d$, respectively. 
    Here, $\lambda = \rbra{\lambda_1, \lambda_2, \dots, \lambda_n} \vdash n$ is a partition of $n$ such that $\lambda_1 \geq \lambda_2 \geq \dots \geq \lambda_n \geq 0$ and $\sum_{i=1}^n \lambda_i = n$, and $\ell\rbra{\lambda}$ is the number of nonzero elements among $\lambda_1, \lambda_2, \dots, \lambda_n$. 
    Weak Schur sampling measures $\rho^{\otimes n}$ and outputs a partition $\lambda \vdash n$. 
    The distribution of the output $\lambda$ is denoted by $\textup{SW}^n_\rho$, with $\textup{SW}^n_\rho\rbra{\lambda} = \tr\rbra{ \Pi_\lambda \rho^{\otimes n} }$,
    where $\Pi_\lambda$ is the projector onto the subspace corresponding to $\mathcal{P}_\lambda \otimes \mathcal{Q}_\lambda^d$.
    The precise formula of $\textup{SW}^n_\rho\rbra{\lambda}$ (see \cite[Proposition 2.23]{OW21}, attributed to \cite{ARS88}) is given as $\textup{SW}^n_\rho\rbra{\lambda} = \dim\rbra{\mathcal{P}_\lambda} \cdot s_\lambda\rbra{p_1, p_2, \dots, p_d}$, if the spectrum of $\rho$ is $p_1 \geq p_2 \geq \dots \geq p_d$, where $s_\lambda\rbra{x_1, x_2, \dots, x_d} = {\det\rbra{x_i^{d+\lambda_j-j}}_{ij}}/{\det\rbra{x_i^{d-j}}_{ij}}$ is the Schur polynomial (see \cite[Theorem 2.13]{OW21} or \cite[Theorem 7.15.1]{Sta10}). 
\end{definition}

Suppose that $\rho$ is a $d$-dimensional quantum state. 
Apply the weak Schur sampling to $\rho^{\otimes m}$ and let $\Lambda = \rbra{\Lambda_1, \Lambda_2, \dots, \Lambda_d} \vdash m$ be the measurement outcome with $\Lambda_1 \geq \Lambda_2 \geq \dots \geq \Lambda_d$ and $\sum_{i=1}^d \Lambda_i = m$. 
Define the estimator
\[
\widehat{S} = -\log\rbra*{\frac{\Lambda_1}{m}}.
\]

\begin{theorem} \label{thm:quantum-upper}
    It holds that $\Pr\sbra{\abs{\widehat{S} - \mathrm{S}_\infty\rbra{\rho}} \leq \varepsilon} \geq 1 - \delta$ for $\varepsilon, \delta \in \rbra{0, 1}$ by setting 
    \[
    m = \ceil*{ \frac{128}{\varepsilon^2}\rbra*{d^2 + d\log\rbra*{\frac{2}{\delta}}} } = O\rbra*{\frac{d^2 + d\log\rbra{1/\delta}}{\varepsilon^2}}.
    \]
\end{theorem}

To prove \cref{thm:quantum-upper}, we need the following properties. 

\begin{proposition} \label{prop:quantum-ineq}
    Let $\mu = \E\sbra{\Lambda_1}$.
    Then, 
    \[
    mp_1 \leq \mu \leq mp_1 + 2\sqrt{2m}.
    \]
    For every $t > 0$, 
    \[
    \Pr\sbra*{ \abs*{\Lambda_1 - \mu} \geq t } \leq 2 \exp\rbra*{-\frac{t^2}{2\rbra{\mu + t/3}}}.
    \]
\end{proposition}

\begin{proof}
    Note that by Greene's theorem \cite{Gre74} (cf.\ \cite[Theorem 2.14]{OW21}), $\Lambda_1$ has the same probability distribution as 
    \[
    Z = \max\cbra{ k : \exists 1 \leq i_1 < i_2 < \dots < i_k \leq m, ~ W_{i_1} \leq W_{i_2} \leq \dots \leq W_{i_k} },
    \]
    the longest nondecreasing subsequence length of $W_1, W_2, \dots, W_m$, 
    where $W_1, W_2, \dots, W_m$ are independent random variables with $\Pr\sbra{W_i = j} = p_j$ (cf.\ \cite[Remark 2.24]{OW21}). 
    The lower bound $\mu \geq mp_1$ follows from $Z \geq \sum_{i=1}^m \mathbf{1}_{\cbra{W_i = 1}}$. 
    The upper bound $\mu \leq mp_1 + 2\sqrt{2m}$ is given by \cite[Lemma 5.1]{OW16}. 
    By \cite[Theorem 3.1]{BLM00}, for every $t > 0$, 
    \[
    \Pr\sbra{Z \geq \mu + t} \leq \exp\rbra*{-\frac{t^2}{2\rbra{\mu + t/3}}}, \qquad \Pr\sbra{Z \leq \mu - t} \leq \exp\rbra*{-\frac{t^2}{2\mu}},
    \]
    which yields the proof.
\end{proof}

We can now give a simple proof of \cref{thm:quantum-upper}.

\begin{proof}[Proof of \cref{thm:quantum-upper}]
    Note that $p_1 \geq 1/d$. 
    Taking $t = \varepsilon m p_1 / 4$ in \cref{prop:quantum-ineq} gives that 
    \[
    \Pr\sbra*{ \abs*{ \frac{\Lambda_1}{m} - p_1 } > \frac{\varepsilon p_1}{2} } \leq \Pr\sbra*{ \abs*{\Lambda_1 - \mu} > \frac{\varepsilon m p_1}{4} } \leq 2 \exp\rbra*{ - \frac{\rbra{\varepsilon mp_1/4}^2}{2\rbra{\mu + \varepsilon mp_1/12}} } \leq \delta,
    \]
    which yields that $\Pr\sbra{ \abs{\widehat{S} - \mathrm{S}_\infty\rbra{\rho}} \leq \varepsilon } \geq 1 - \delta$. 
\end{proof}

\subsection{Lower Bounds}

\begin{theorem}\label{thm:quantum-lower}
    Any estimator that estimates the min-entropy of a $d$-dimensional quantum state $\rho$ to within additive error $\varepsilon$ with success probability at least $1 - \delta$ requires sample complexity at least 
    \[
    \Omega\rbra*{\frac{d^2 + d\log\rbra{1/\delta}}{\varepsilon^2}}.
    \]
\end{theorem}

To prove \cref{thm:quantum-lower}, we need a technical lemma from \cite{Wan26}. 

\begin{lemma}[Special case of {\cite[Lemma 3.2]{Wan26}}] \label{lemma:wan26}
    Let $\nu_0$ and $\nu_1$ be two distributions on $\sbra{-1, D-1} \cap \sbra{-B, B}$. 
    Define
    \[
    \sigma_b^{\rbra{n}} = \E_{\substack{Z \sim \nu_b \\ \ket{u} \sim \textup{Haar}\rbra{\mathbb{C}^D}}} \sbra*{ \varsigma_{Z, u}^{\otimes n} }, \qquad \varsigma_{Z, u} = \frac{1+Z}{D} \ketbra{u}{u} + \frac{1-Z/\rbra{D-1}}{D} \rbra{\mathbb{I} - \ketbra{u}{u}}.
    \]
    If $R \coloneqq 2nB^2/\rbra{D-1}^2 < 1$, then 
    \[
    \frac{1}{2} \Abs*{ \sigma_0^{\rbra{n}} - \sigma_1^{\rbra{n}} }_1 \leq \sqrt{\frac{R}{1-R}}.
    \]
\end{lemma}

We can now give a simple proof of \cref{thm:quantum-lower}.

\begin{proof}[Proof of \cref{thm:quantum-lower}]
    Consider the quantum state
    \[
    \rho_v = \frac{\mathbb{I}}{d} + \frac{4\varepsilon}{d-1} \rbra*{ \ketbra{v}{v} - \frac{\mathbb{I}}{d} } = \frac{1+4\varepsilon}{d} \ketbra{v}{v} + \frac{1-4\varepsilon/\rbra{d-1}}{d} \rbra*{ \mathbb{I} - \ketbra{v}{v} },
    \]
    where $\ket{v}$ is a Haar-random unit vector. 
    Note that the spectrum of $\rho_v$ is the probability distribution $q^{\rbra{1}}$ used in \cref{thm:clower}. 
    Assume that there is an estimator with sample complexity $m$ that estimates the min-entropy of a quantum state to within additive error $\varepsilon$ with success probability at least $2/3$, it implies a tester that distinguishes $\rbra{\mathbb{I}/d}^{\otimes m}$ from $\E_{v}\sbra{\rho_v^{\otimes m}}$ with success probability at least $2/3$. 
    The Helstrom--Holevo bound \cite{Hel67,Hol73} gives that 
    \begin{equation} \label{eq:qlb}
        \frac{1}{2} \Abs*{ \rbra*{\frac{\mathbb{I}}{d}}^{\otimes m} - \E_{v \sim \textup{Haar}\rbra{\mathbb{C}^d}} \sbra*{\rho_v^{\otimes m}}}_1 \geq \frac{1}{3}.
    \end{equation}
    On the other hand, \cref{lemma:wan26}, with $D = d$, $B = 4\varepsilon$, $n = m$, $L = 0$, $\nu_0$ being the distribution supported on $\cbra{0}$, and $\nu_1$ being the distribution supported on $\cbra{4\varepsilon}$, gives that if $32m\varepsilon^2 < \rbra{d-1}^2$,
    \begin{equation} \label{eq:qub}
    \frac{1}{2} \Abs*{ \rbra*{\frac{\mathbb{I}}{d}}^{\otimes m} - \E_{v \sim \textup{Haar}\rbra{\mathbb{C}^d}} \sbra*{\rho_v^{\otimes m}}}_1 \leq \sqrt{\frac{32m\varepsilon^2}{\rbra{d-1}^2 - 32m\varepsilon^2}}.
    \end{equation}
    Combining \cref{eq:qlb,eq:qub} gives 
    \[
    m \geq \frac{\rbra{d-1}^2}{512\varepsilon^2} = \Omega\rbra*{\frac{d^2}{\varepsilon^2}}.
    \]
    Otherwise, the right hand side of \cref{eq:qub} is less than $1/3$, contradicting \cref{eq:qlb}. 
    The proof is completed by combining the classical lower bound in \cref{thm:clower}.
\end{proof}

\section*{Acknowledgment}

The author used Large Language Models to explore proof strategies. 
The author then distilled and organized the arguments into the short and simple proofs presented in this paper.
The manuscript was written by the author, who assumes full responsibility for its content.

\addcontentsline{toc}{section}{References}

\bibliographystyle{alphaurl}
\bibliography{main}

\appendix

\end{document}